\documentclass[twocolumn,nofootinbib,amsmath]{revtex4}
\usepackage{graphicx}

\begin{document}

\title{Vacuum  Choices and  the Predictions  of Inflation}  \author{C. 
  Armend\'ariz-Pic\'on}               \email{armen@oddjob.uchicago.edu}
\affiliation{Enrico  Fermi Institute and  Department of  Astronomy and
  Astrophysics,  \\   University  of  Chicago.}    \author{Eugene  A.  
  Lim}\email{elim@jasmine.uchicago.edu} \affiliation{Center for
  Cosmological Physics and  Department of Astronomy  and Astrophysics, \\
  University of Chicago.}

\begin{abstract}
  In the presence of a short-distance cutoff, the choice of a vacuum
  state in an inflating, non-de Sitter universe is unavoidably
  ambiguous.  The ambiguity is related to the time at which initial
  conditions for the mode functions are specified and to the way the
  expansion of the universe affects those initial conditions.  In this
  paper we study the imprint of these uncertainties on the predictions
  of inflation.  We parametrize the most general set of possible
  vacuum initial conditions by two phenomenological variables.  We
  find that the generated power spectrum receives oscillatory
  corrections whose amplitude is proportional to the Hubble parameter
  over the cutoff scale.  In order to further constrain the
  phenomenological parameters that characterize the vacuum definition,
  we study gravitational particle production during different
  cosmological epochs.
\end{abstract}

\maketitle

\section{Introduction}
Inflation   \cite{Linde}   causally  explains   the   origin  of   the
superhorizon density perturbations that seed the structures we observe
in the  universe.  During  inflation, the physical  size of  any given
perturbation mode grows faster than the Hubble radius.  In particular,
observationally relevant  modes start  well within the  Hubble radius,
cross  it and  ``freeze''  till  they reenter  the  Hubble radius  and
develop into galaxies,  clusters, etc. \cite{MuFeBr}. Causality within
the  Hubble radius  naturally  allows us  to  determine their  initial
amplitude  by postulating  that  perturbations start  in their  vacuum
state.

In Minkowski space there  exists a well-defined, unique, vacuum state,
but in any expanding universe---and in particular, during inflation---
the  notion of  a  vacuum is  ambiguous  \cite{BirrellDavies}.  In  de
Sitter space  there exists a  concrete set of vacuum  states invariant
under  the   symmetry  group  of  the  spacetime.    Such  vacua,  the
Bunch-Davies vacuum  and the $\alpha$-vacua,  have received widespread
attention  in the recent  literature \cite{vacua}.   However, strictly
speaking, an inflating spacetime is \emph{not} de Sitter spacetime. de
Sitter  space admits  a time-like  Killing vector  field,  whereas the
non-existence of a time-like Killing  vector field is what singles out
cosmological spacetimes.   A Friedmann-Robertson-Walker universe whose
scale factor grows quadratically in cosmic time ($a\propto t^2$) is an
inflating  spacetime,  and  it  is  far from  obvious  why  de  Sitter
spacetime, where $a$ grows exponentially with time ($a\propto e^{Ht}$)
shall be a good description of such a spacetime.  More importantly, in
de Sitter space  scalar cosmological perturbations are ``ill-defined''
\cite{DeruelleMukhanov}, as signaled for instance by the divergence of
the  power spectrum  in  the  de Sitter  limit.   Thus, during  cosmic
inflation one still has to face the problem of defining a vacuum state
in an expanding universe that does not admit any particular additional
symmetry.

Recent advances in  the formulation of quantum field  theory in curved
spacetimes  \cite{Wald} suggest  that  such a  preferred vacuum  state
might   not   exist.   Whereas   one   can  construct   mathematically
well-defined quantum field-theories in globally hyperbolic spacetimes,
they do not single out any particular quantum state.  Allowed physical
states  only  have to  satisfy  the  Hadamard  condition, which  is  a
condition  on the  ultraviolet ($k\to  \infty$) behavior  of two-point
functions.  In this work we  assume physics to be unknown above cutoff
energies  $\Lambda$.  The  presence  of this  ultraviolet cutoff  thus
renders the Hadamard condition  inapplicable.  Therefore, i) in curved
spacetimes there  is no  single preferred quantum  state and  ii) even
though there  is a class  of preferred states,  in the presence  of an
ultraviolet  cutoff $\Lambda$  the  condition that  singles out  those
states is not applicable.

Nevertheless,  it   is  still  possible   to  single  out  a   set  of
``reasonable''  vacuum  states by  appealing  to  the  flatness of  an
expanding universe at short distances. In the limit where the physical
length  $\lambda$ of  a mode  is  infinitely smaller  than the  Hubble
radius $H^{-1}$,  the expansion is  completely negligible.  Therefore,
in  the  conventional  treatments  of  inflation,  the  vacuum  for  a
particular mode  is chosen when $\lambda/H^{-1}\to 0$,  i.e.  when the
vacuum state  agrees with the Minkowski vacuum.   However, if physical
laws are unknown below certain cutoff length $\Lambda^{-1}$, the limit
$\lambda\to  0$   leads  into  a  region  where   physics  is  unknown
\cite{Brandenberger}.   In  the presence  of  a  cutoff,  in order  to
minimize the ambiguity  related to the expansion of  the universe, and
in order to  avoid the region of unknown physics, the  best one can do
is to define  the vacuum at the time the  physical length $\lambda$ of
the  mode   equals  the  fundamental  length  scale   of  the  theory,
$\Lambda^{-1}$  \cite{Jacobson}.   In   that  case,  the  conventional
predictions of inflation get modified due to the finite effects of the
expansion of the universe \cite{Danielsson}.  These corrections can be
expanded as a  series in $H/\Lambda$.  Let us  stress though, that the
corrections   we  are   talking   about  have   little   to  do   with
trans-Planckian  physics  (see \cite{Brandenberger2}  for  a review).  
Rather, they arise from the uncertainties related to the definition of
vacuum in  an expanding universe.  If  we knew how  to uniquely define
the vacuum of a field in an expanding universe, there would not be any
uncertainty at all.  In fact, in  the limit of no expansion, $H\to 0$,
all vacuum prescriptions we consider here agree, i.e.  to zeroth order
in  $H/\Lambda$ all  of  them yield  the  same ``conventional''  power
spectrum prediction.

The  precise nature  of the  small  $H/\Lambda$ corrections  is not  a
question  of academic  interest  only.  In  some inflationary  models,
$H/\Lambda$ might be  large enough in order for  linear corrections to
leave   an    \emph{observable}   imprint   on    the   CMB   spectrum
\cite{BergstromDanielsson,LimOkamoto},  whereas  quadratic corrections
are  expected  to  be  unobservable  \cite{KKLS}.   In  addition,  the
definition of a  vacuum also directly affects the  number of particles
produced  due   to  the  expansion  of  the   universe.   By  imposing
observational  constraints  on   the  amount  of  the  gravitationally
produced particles  one can thus  gain information about  the realized
vacuum \cite{BirrellDavies, Starobinsky}.

The goal  of this paper  is not to  assess which vacuum choice  is the
correct one, as we have argued  that there might be no answer for that
question.  Our goal is rather  to take a phenomenological approach and
find out  what is a reasonable set  of possible vacua, and  how and to
what  extent these  possibly  different vacua  alter the  conventional
predictions  of inflation.   Then, instead  of relying  on theoretical
arguments to single out the vacuum that was realized during inflation,
we shall rely on observations  to put constraints on possibly realized
vacua.

\section{Formalism}     
For completeness, we  shortly review in this section  the formalism of
the generation of perturbations  in an inflating spacetime. The reader
might want to  skip to the next section and  eventually refer back for
notational  details.    We  also  summarize  our   notation  in  Table
\ref{table:notation}.

We consider in  the following power-law inflation in  a spatially flat
FRW    universe,    
\begin{equation}\label{eq:metric}
  ds^2=a^2(\eta)(d\eta^2-d\vec{x}^2).
\end{equation}
Power-law  inflation  is  simple  enough to  allow  a  straightforward
treatment of the  equations of motion of the  perturbations, and it is
general enough  to accommodate a wide realistic  class of inflationary
behaviors.  During power law inflation the scale factor is given by
\begin{equation}\label{eq:scale-factor}
a\propto|\eta|^{\frac{p}{1-p}},
\end{equation}
where $p>1$.  In the limit  $p\to\infty$ one recovers de Sitter space,
$a\propto  -1/\eta$.   Note that  conformal  time  is negative  during
inflation ($p>1$).   Equation (\ref{eq:scale-factor}) also  applies to
any stage of power-law expansion ($0\leq p<1$), when conformal time is
positive.  For $p=0$ one recovers Minkowski space.

The  behavior  of scalar  and  tensor  perturbations during  inflation
driven by a single scalar field  can be described in terms of a single
scalar variable $v(\eta,\vec{x})$ \cite{MuFeBr, GarrigaMukhanov}.  For
scalar perturbations  the variable $v$ is a  particular combination of
metric and scalar  field perturbations, whereas in the  case of tensor
perturbations,  the  variable  $v$   is  simply  proportional  to  the
amplitude  of  the  gravitational  waves.   The dynamics  of  $v$  are
determined    by   the   quadratic    Lagrangian\footnote{For   scalar
  perturbations the  quoted Lagrangian only applies during  a stage of
  power-law inflation.  See \cite{MuFeBr}, Eq. (11.1), for the correct
  expression in an arbitrary background.}
\begin{equation}\label{eq:action}
  L=\frac{1}{2}\int d^3x\left[v'^2-\delta^{ij}
        \frac{\partial v}{\partial x^i}\frac{\partial v}{\partial x^j}
        +\frac{a''}{a}v^2\right],
\end{equation}
where a prime denotes a  derivative with respect to conformal time and
$i,j$ run from $1$ to $3$.

The  classical variable $v$  can be  quantized following  the standard
rules \cite{MuFeBr}.   Upon quantization,  $v$ turns into  an operator
$\hat{v}$,  which  can  be  expanded in Fourier modes, 
\begin{equation}\label{eq:mode-expansion}
  \hat{v}=\frac{1}{\sqrt{2}}\int \frac{d^3 k}{(2 \pi)^{3/2}}
  \left(v_k  (\eta) e^{i\vec{k}\cdot\vec{x}}  \hat{a}_k
    +v_k^*(\eta)  e^{-i \vec{k}\cdot\vec{x}}\hat{a}^\dag_k\right).
\end{equation}
The mode functions $v_k(\eta)$ obey the differential equation
\begin{equation}\label{eq:motion}
v_k''+\left(k^2-\frac{a''}{a}\right) v_k=0,
\end{equation}
and the operators $a_k$  can be interpreted as annihilation operators,
$[a_k,a_{k'}^\dag]=\delta(\vec{k}-\vec{k'})$, if the $v_k$ satisfy the
normalization condition
\begin{equation}\label{eq:normalization}
v'_kv^*_k-v'^*_k v_k=-2i.
\end{equation}
In terms  of creation and  annihilation operators, the  Hamiltonian of
the system (\ref{eq:action}) is given by
\begin{eqnarray}\label{eq:Hamiltonian}
\hat{\mathcal{H}}=\frac{1}{4}\int  d^3 k\,
\big[({v'_k}^2+\omega_k^2 v_k^2)\, a_k a_{-k}+ \\ \nonumber
+(|v_k'|^2+\omega_k^2 |v_k|^2)a_k^\dag a_k+ h.c.\big],
\end{eqnarray}
where we have defined the squared frequency
\begin{equation}\label{eq:frequency}
  \omega_k^2=k^2-\frac{a''}{a}.
\end{equation}

Given a mode  expansion, i.e.  given a particular  set of annihilation
operators  $a_k$, the vacuum  of the  field $|0\rangle$  is implicitly
defined by  the condition $a_k|0\rangle=0$.  Since, in  the absence of
further requirements  Eq. (\ref{eq:motion}) does not  suffice alone to
uniquely   determine  the  mode   functions  $v_k$,   infinitely  many
definitions of vacuum are  possible \cite{BirrellDavies}.  In the next
section  we shall  formulate  additional criteria  that constrain  the
possible vacuum state choices.

Once a vacuum  state has been determined, it is  possible to study the
imprint of the vacuum  upon observable quantities, such as temperature
anisotropies   in  the  cosmic   microwave  background.    For  scalar
perturbations, the amplitude of such fluctuations can be characterized
by the power spectrum of the Bardeen variable \cite{MuFeBr}
\begin{equation}\label{eq:Bardeen}
 \zeta=\frac{\sqrt{4\pi\, p}}{M_{Pl}}\frac{v}{a},
\end{equation}
where  $M_{Pl}$  is  the  Planck mass  $M_{Pl}^2=G^{-1}$.   The  power
spectrum $\mathcal{P}$  is then implicitly defined  by the correlation
function of the $\zeta$ variable,
\begin{equation}\label{eq:power-def}
  \langle 0|\hat{\zeta}^\dag(\eta,\vec{x})
  \hat{\zeta}(\eta,\vec{x}+\vec{r})|0\rangle
  \equiv \int  \frac{dk}{k}   \frac{\sin(kr)}{kr} \,\mathcal{P}_k.
\end{equation}
Substituting  the mode  expansion  (\ref{eq:mode-expansion}) into  the
last definitions one arrives at  the following expression for the power
spectrum,
\begin{equation}\label{eq:power-exp}
  \mathcal{P}_k=\frac{p}{\pi M_{Pl}^2}\frac{k^3|v_k|^2}{a^2}.
\end{equation}

The value of  the power spectrum for  a given $k$ is a  measure of the
mean  square fluctuations  of  the variable  $v$ over  \emph{comoving}
distances $r\approx 1/k$ \cite{MuFeBr}.  Hence, the physical extent of
a field fluctuation labeled by $k$ is
\begin{equation}\label{eq:length}
  \lambda=\frac{a}{k},
\end{equation} 
which depends on time for fixed $k$. In an expanding universe there is
a natural physical distance scale given by the Hubble radius $H^{-1}$,
where $H\equiv a'/a^2$ is the Hubble parameter.  We denote by $\theta$
the  (dimensionless) ratio of  the physical  distance associated  to a
given mode $k$ and the Hubble radius,
\begin{equation}\label{eq:theta}
  \theta\equiv\frac{\lambda}{H^{-1}}=\frac{p}{1-p}\frac{1}{k\eta}.
\end{equation}
In the limit of no expansion, $p\to 0$, $\theta$ tends to zero.  The
parameter  $\theta$  will  play   a  crucial  role  in  our  further
discussions.

\begin{table}
\caption{\label{table:notation}Summary of notation}
\begin{ruledtabular}
\begin{tabular}{l|l|r}
Symbol & Meaning & Equation \\
\hline
$\beta$ and $\alpha$& Bogolubov coefficients & (\ref{eq:transformation})\\
$\zeta$ & Bardeen variable& (\ref{eq:Bardeen})\\
$\eta$ & Conformal time &  (\ref{eq:metric}) \\
$\theta$ & Mode length over Hubble radius& (\ref{eq:theta})\\
$\Lambda$ & High energy cutoff & \\
$\lambda$ & Physical size of a mode& (\ref{eq:length})\\
$\nu$ & Index of Hankel function & (\ref{eq:nu}) \\
$a$ & Scale factor & (\ref{eq:metric}) \\
$k$ & Labels a perturbation mode & \\
$\mathcal{P}$ & Power spectrum & (\ref{eq:power-def}) \\
$p$ & Power-law expansion exponent & (\ref{eq:scale-factor}) \\ 
$v$ & The quantization variable & (\ref{eq:action}) \\ 
$v_k$ & Mode function & (\ref{eq:mode-expansion}) \\
$X$ and $Y$  & Vacuum parameters & (\ref{eq:prescription}) \\
Subscript 0 & Initial time & \\
$'$ & $d/d\eta$ & \\
\end{tabular}
\end{ruledtabular}
\end{table}

\section{Vacuum     Choice}
Several vacuum  choices have been  proposed in the literature,  and we
discuss some of  them in the Appendix. Here we  shall pursue however a
different approach.  Instead of declaring one of the several different
vacua to  be the correct one,  we shall abstract the  property all the
different vacua share and use  that property to define a general class
of ``sensible'' vacuum states.

Specifying  a  set of  mode  functions  $v_k(\eta)$  is tantamount  to
choosing a  vacuum state $|0\rangle$.  Because $v_k$  is determined by
the equation of  motion (\ref{eq:motion}), all that has  to be done to
define a vacuum is to  specify initial conditions for $v_k$.  We shall
require  of any  vacuum state  that in  the limit  where  the physical
length  of the  perturbation is  much smaller  than the  Hubble radius
($\theta\ll  1$), the mode  functions approach  the ones  in Minkowski
space $v_k\approx e^{-i  k \eta}/\sqrt{k}$.  In fact, in  such a limit
cosmic expansion should be irrelevant. In other words, when $\theta\to
0$ the values  of $v_k$ and its time  derivative respectively approach
(up to an irrelevant phase)
\begin{equation}\label{eq:Minkowski}
v_k=\frac{1}{\sqrt{k}}\quad \text{and} \quad
v'_k=-i\sqrt{k}.
\end{equation}
All the vacuum choices we discuss in the appendix share this property.
However, in  an inflating universe (where $H$  remains finite) physics
is not  well defined in  the limit $\theta\to  0$.  In that  limit the
physical  length of the  perturbation is  infinitely smaller  than the
Planck   length.   Hence,   one  cannot   rely  on   our  conventional
understanding  of  physics  in  that regime\footnote{In  Pre-Big  Bang
  \cite{GasperiniVeneziano}     and    Epkyrotic/Cyclic    \cite{EpCy}
  scenarios,  the  universe contracts  from  Minkowski  space at  past
  infinity.  In that  case the limit $\theta\to 0$  is physically well
  defined.  For  those models, the  ambiguities we are  considering in
  this work  can be avoided.}.  Thus,  for a given mode  $k$, the best
one can  do is pick  a vacuum by  prescribing the values of  $v_k$ and
$v'_k$ at a finite time $\eta_0$  such that the physical length of the
mode  is  much  smaller than  the  Hubble  radius  (in order  for  the
expansion  to be  ``unimportant''), but  much larger  than  the Planck
scale  (in order for  trans-Planckian effects  to be  negligible).  We
shall take  $\eta_0$ to be  the time when  the physical length  of the
mode equals a given (fixed) length scale $\Lambda^{-1}$,
\begin{equation}\label{eq:crossing}
  \frac{a(\eta_0)}{k}=\Lambda^{-1}.
\end{equation}
The scale $\Lambda$  is the highest possible scale  where we can trust
our  understanding  of physics.   Conventionally  it  is assumed  that
$\Lambda$  is the  Planck scale  $\Lambda=M_{Pl}\approx  10^{19}$ GeV,
although  the  cutoff  could  be  as low  as  $\Lambda\approx  1$  TeV
\cite{AADD}.   Note  that the  time  of  cutoff  crossing $\eta_0$  is
$k$-dependent.

According  to Eq.   (\ref{eq:crossing}), initial  conditions  for each
mode  are  prescribed at  the  same  physical  length $\Lambda^{-1}$.  
Hence,  the  hypersurface  where  initial  conditions  are  chosen  is
timelike.  For  comparison, it  is going to  be useful to  consider an
additional  hypersurface where  initial conditions  for the  modes are
specified.  We shall choose this  hypersurface to be the constant time
spatial  section   where  the  Hubble  parameter   equals  the  cutoff
$\Lambda$,
\begin{equation}\label{eq:equal-time}
  H(\eta_0)=\Lambda.
\end{equation}
In this case $\eta_0$ does not depend on $k$.  Note that in the latter
case, observationally  relevant modes might be  trans-Planckian at the
time initial conditions are fixed \cite{Brandenberger}.

Because  initial conditions  for the  mode  functions are  fixed at  a
finite time  rather that at $\eta_0=-\infty$,  one expects corrections
to Eqs.   (\ref{eq:Minkowski}) due to  the expansion of the  universe. 
The only dimensionless quantity one  can construct from the two scales
that appear to be relevant in the problem---the physical length of the
mode  $\lambda$  and  the  Hubble  radius  $H^{-1}$---is  their  ratio
$\theta$.  Hence, those  corrections are expected to be  a function of
$\theta_0$,  the  value of  $\theta$  evaluated  at  the time  initial
conditions are  imposed.  Since  we assume $\theta_0$  to be  small at
that time,  and because those  corrections should vanish in  the limit
$\theta_0\to  0$,  we  can  expand  them  in  a  power  series  around
$\theta_0=0$.  We  shall concentrate  on the lowest  order corrections
(first order), i.e.  we impose \cite{MuFeBr}
\begin{subequations}\label{eq:prescription}
\begin{eqnarray}
  v_k(\eta_0)&=&\frac{e^{i \phi_1}}{\sqrt{k}}
  \left[1+\frac{X+Y}{2}\, \theta_0+
    \mathcal{O}(\theta_0^2)\right] \\
  v_k'(\eta_0)&=&-i\sqrt{k}e^{i \phi_2}
  \left[1+\frac{Y-X}{2}\, \theta_0
    +\mathcal{O}(\theta_0^2)\right]. 
\end{eqnarray}
\end{subequations}
Here,  $Y$  and $X$  are  two  complex  parameters, and  $\phi_1$  and
$\phi_2$ are  two arbitrary real phases.   The normalization condition
(\ref{eq:normalization})  implies  that   the  phases  are  the  same,
$\phi_1=\phi_2$.  Because  the power spectrum  (\ref{eq:power-exp}) is
invariant  under  $v_k\to  e^{i\phi}v_k$  we shall  drop  the  phases,
$\phi_1=\phi_2=0$.   In the  same  way, Eq.   (\ref{eq:normalization})
constraints the values of $Y$,
\begin{eqnarray}\label{eq:constraint}
  \mathrm{Re}(Y)=0. 
\end{eqnarray}
The variables $X$ and $Y$ are two phenomenological parameters that
characterize the choice of vacuum to first order in $\theta_0$. Our
approach could be generalized to higher orders by including additional
parameters, but for our purposes it will suffice to consider the
lowest order corrections.  All the sensible vacuum prescriptions we
are aware of can be cast in the form (\ref{eq:prescription}).  In
fact, our point is that any vacuum prescription that can be cast in
the form (\ref{eq:prescription}) is a sensible vacuum choice.  The
specific values of $X$ and $Y$ for the particular vacuum prescriptions
that have been considered in the literature are listed in Table
\ref{table:vacua} (see the Appendix for details).  If $X$ and $Y$ were
$p$-independent, one could use the information about vacua in
spacetimes other than Minkowski to restrict the values of $X$ and $Y$.
For instance, if it turned out that the Bunch-Davies vacuum is the
only consistent vacuum in de Sitter spacetime \cite{vacua}, one could
directly compute the values of $X$ and $Y$ for de Sitter and apply
them to non-de Sitter spacetimes.  However, at a phenomenological level
$X$ and $Y$ could be $p$-dependent, as happens for instance in the
``conventional'' vacuum prescription.

\begin{table}
\caption{\label{table:vacua}Properties of different vacuum prescriptions}
\begin{ruledtabular}
\begin{tabular}{l|c|c}
Vacuum prescription & $X$ & $Y$ \\
\hline
Conventional & $0$  &$i (1-2p)/(1-p)$ \\ 
Adiabatic $\geq$ 1st order & 0 & 0 \\
Hamiltonian diagonalization & 0 & 0  \\
Danielsson & $-i$ & $i$ \\
\end{tabular}
\end{ruledtabular}
\end{table}

Our next task is  to implement the ``generalized'' vacuum prescription
(\ref{eq:prescription}) during  a stage of power-law  inflation.  In a
power-law  expanding, not necessarily  inflating universe  the general
solution of Eq.  (\ref{eq:motion}) is
\begin{equation}\label{eq:solution}
v_k(\eta)=|\eta|^{1/2}\left[A_k H_\nu(|k \eta|)
  +B_k H^*_\nu(|k\eta|)\right],
\end{equation} 
where the  $H_\nu$ is  the Hankel function  \cite{AbramowitzStegun} of
the first kind if conformal time is negative ($p>1$) and of the second
kind if conformal  time is positive ($p<1$). The  index $\nu$ is given
by
\begin{equation}\label{eq:nu}
  \nu=\frac{3}{2}+\frac{1}{p-1}.
\end{equation}

We  shall  determine  the   values  of  the  time-independent  complex
coefficients   $A_k$   and    $B_k$   by   imposing   the   conditions
(\ref{eq:prescription})  on the  exact  solution (\ref{eq:solution}).  
For  large values  of $|k\eta|$  the Hankel  function $H_\nu$  has the
expansion
\begin{eqnarray}\label{eq:expansion}
H_\nu(|k\eta|)\approx \sqrt{\frac{2}{\pi|k\eta|}}
\left[1-i \frac{4\nu^2-1}{8k\eta}+\mathcal{O}(\theta^2)\right]\times \\
 \times\exp\left[-ik\eta\mp i 
   \pi\left(\frac{\nu}{2}+\frac{1}{4}\right)\right], 
 \nonumber
\end{eqnarray}
which is  valid for both  $p>1$ (upper sign)  and $p<1$ (lower  sign). 
Plugging  the expansion (\ref{eq:expansion})  into (\ref{eq:solution})
and matching to Eqs.   (\ref{eq:prescription}) we find (to first order
in $\theta_0$)
\begin{subequations}\label{eq:AandB}
\begin{eqnarray}
  A_k&=&\sqrt{\frac{\pi}{2}}
  \left[1+Y\,\frac{\theta_0}{2}
    -\frac{i}{2}\,\frac{1-2p}{1-p}\theta_0  \right]
  e^{i\varphi}, \\
  B_k&=&\sqrt{\frac{\pi}{2}}\,
  X\,\frac{\theta_0}{2} e^{-i\varphi}, \\
  \varphi&=&\frac{p}{1-p}\,\theta_0^{-1}
  \pm \frac{\pi}{2} \left(\frac{1-2p}{1-p}\right).
\end{eqnarray}
\end{subequations}
For  $\theta_0=0$,  $A_k=\sqrt{\pi/2}$  and  $B_k=0$.  These  are  the
values one conventionally  chooses when computing inflationary spectra
\cite{MuFeBr}. For finite $\theta_0$  one hence obtains corrections to
these ``conventional'' results.

Let us emphasize again that the corrections in powers of $\theta_0$ we
consider  here are not  directly related  to trans-Planckian  effects. 
Our corrections  become large on length  scales of the  Hubble radius. 
Their  origin  can be  ultimately  traced  back  to the  ambiguity  in
defining the notion  of a particle in an  expanding universe, when the
Compton  wavelength  has  a  size  comparable  to  the  Hubble  radius
\cite{BirrellDavies}.  Unknown trans-Planckian  physics only enter our
discussion  by preventing us  from taking  the limit  $\theta_0\to 0$,
making those corrections finite (but still small) rather than zero.

\section{Imprint on the Power Spectrum}
The power spectrum (\ref{eq:power-def})  is defined through the vacuum
expectation value  of the two-point  function of the  Bardeen variable
$\zeta$.   Hence, the  choice  of vacuum  directly  affects the  power
spectrum.

Substituting   the  mode   function  (\ref{eq:solution})   into   Eq.  
(\ref{eq:power-exp}),   using   the   values   of   the   coefficients
(\ref{eq:AandB}) and taking Eq. (\ref{eq:constraint}) into account one
easily finds in the long-wavelength limit ($k\eta\ll1$)
\begin{eqnarray}\label{eq:corrected-power}
\mathcal{P}=\mathcal{P}_C\Big[1-\mathrm{Re}(X)\,\theta_0
\cos\left(\frac{2 p\, \theta_0^{-1}+\pi\,(1-2p)}{1-p}\right)- 
\nonumber \\ 
{}-\mathrm{Im}(X)\,\theta_0
\sin\left(\frac{2 p\, \theta_0^{-1}+\pi\,(1-2p)}{1-p}\right)+
\mathcal{O}(\theta_0^2)\Big],
\end{eqnarray} 
where $\mathcal{P}_C$ is the conventional power spectrum generated
during power-law inflation,
\begin{eqnarray}
  \mathcal{P}_C=\left[\frac{|\Gamma(\nu)|^2}{\pi^2} 
    \left(\frac{2(p-1)}{p}\right)^{\frac{2p}{p-1}}
      \left(\frac{H_*}{\Lambda}\right)^{\frac{2}{p-1}}\right]\times 
    \nonumber\\
\times\,  p \frac{H_*^2}{M_{Pl}^2}\left(\frac{k}{k_*}\right)^{-\frac{2}{p-1}}.
\end{eqnarray}
In the last  formula, $H_*$ denotes the value  of the Hubble parameter
when the particular reference  mode crosses the cutoff $\Lambda$.  The
spectral index  $n_S$ of the  conventional power spectrum  is directly
related to $p$ by the equation
\begin{equation}
n_S=1-\frac{2}{p-1}.
\end{equation}  
Current observations \cite{MAP,InflMAP} set the lower limit $n_S>0.9$,
which  implies $p>21$.   Notice that  in the  limit  $p\to\infty$, the
conventional power spectrum diverges, $\mathcal{P}_C\propto p$.  In de
Sitter space  the theory of  scalar cosmological perturbations  is not
well-defined \cite{DeruelleMukhanov}.

The value of $\theta_0$ depends on the hypersurface at which initial
conditions are chosen. If the initial time is chosen to be
cutoff-crossing, Eq. (\ref{eq:crossing}), $\theta_0$ is given by
\begin{equation}\label{eq:theta-l}
  \theta_0=\frac{H_*}{\Lambda}\left(\frac{k}{k_*}\right)^{-1/p}.
\end{equation}
On the other hand, if the time is chosen to be the same for all modes,
Eq.  (\ref{eq:equal-time}), $\theta_0$ is given by
\begin{equation}\label{eq:theta-t}
  \theta_0=\left(\frac{H_*}{\Lambda}\right)^p
  \frac{k_*}{k}.
\end{equation}
In  both formulas  $H_*$  is again  the value  of  $H$ at  the time  a
comoving reference mode $k_*$ crosses the cutoff.

The  term in  the  square bracket  of Eq.   (\ref{eq:corrected-power})
contains the corrections to  the standard predictions of inflation due
to the ambiguity in the choice of vacuum.  For arbitrary values of $X$
these  corrections are oscillatory,  with amplitude  $\theta_0$. First
order $\theta_0$ corrections are absent if and only if
\begin{equation}\label{eq:no-linear}
X=0.
\end{equation}
Taking   into  account   the  constraint   (\ref{eq:constraint}),  our
generalized vacuum choice spans a 3-dimensional space parametrized by,
say,        $\mathrm{Im}(Y)$,        $\mathrm{Re}(X)$       and
$\mathrm{Im}(X)$.   Equation (\ref{eq:no-linear}) just  says that
the set of  parameters for which there are no  linear corrections is a
two-dimensional plane, i.e.  it is of zero measure in parameter space.

\begin{figure}
\begin{center}
  \includegraphics[width=8cm]{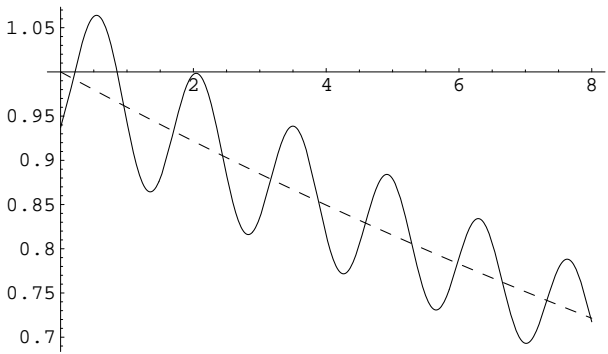}
\caption{A plot---in arbitrary units---of the conventional (dashed) 
  and  the corrected  power  spectra (solid)  vs. $\log(k/k_*)$.   The
  values  of  the  parameters  are  $p=50$  (uncorrected  $n_S=0.96$),
  $H_*/\Lambda=10^{-2}$ and $\theta=-4+8 i$.\label{fig:power}}
\end{center}
\end{figure}

If the  vacuum is chosen at cutoff  crossing, Eq.  (\ref{eq:theta-l}),
corrections   are   hence    generically   linear   in   $H_*/\Lambda$
\cite{Danielsson,EGKS}.  For large values of $p$, the frequency of the
oscillations $\omega_{osc}$ [as a function of $\log(k/k_*)$] is mildly
$k$-dependent.  To lowest order in $1/p$, it is given by
\begin{equation}
\omega_{osc}=\frac{2}{p-1}\left(\frac{H_*}{\Lambda}\right)^{-1}.
\end{equation}
In the de  Sitter limit, $p\to \infty$, there  are no oscillations.  A
plot  of both  the conventional  and the  corrected power  spectrum is
shown in Figure \ref{fig:power}.

If the  vacuum for  each mode is  chosen at  the time when  the Hubble
parameter equals the cutoff,  Eq.  (\ref{eq:theta-t}) the amplitude of
the  corrections  is  proportional to  $(H_*/\Lambda)^p$.   Therefore,
unless $H_*\approx  \Lambda$ those corrections  are highly suppressed,
since for reasonable values of the spectral index, $p$ is large.

In the Appendix we discuss  many of the vacuum prescriptions that have
been proposed in  the literature and how they  translate into specific
values  of  $X$  and  $Y$.    The  results  are  summarized  in  Table
\ref{table:vacua}.  Inspection  of that table shows that  only for the
Danielsson prescription  corrections to the power  spectrum are linear
in $H/\Lambda$.  For the remaining prescriptions, the adiabatic vacuum
(of   order   bigger  than   one)   and  Hamiltonian   diagonalization
\cite{NiPaCa,BoGiVe},  the lowest order  corrections are  \emph{at the
  most}  quadratic\footnote{This  conclusion   is  in  agreement  with
  \cite{NiPaCa}, although the adiabatic  vacuum we discuss here is not
  the ``adiabatic vacuum'' considered by those authors.}.  Let us note
however that  in our opinion,  no strong theoretical  argument singles
out any of the different vacuum choices.

\section{Particle Production}
Given  the ambiguities  in  the definition  of  a vacuum  state in  an
expanding  universe, one  might  take a  phenomenological approach  to
constrain plausible  vacua.  In general,  one expects particles  to be
produced  due to  the  changing gravitational  field  in an  expanding
universe.  The  amount of  particles produced depends  on the  way the
vacuum is defined. Hence, by requiring the rate of particle production
to  be  negligible, one  can  put  constraints  on the  vacuum  choice
\cite{Starobinsky}.

Consider  a graviton  (or a  massless scalar  field) propagating  in a
fixed  expanding  universe.  The  amplitude  $h$  of  any of  the  two
polarization  states  of  the  graviton  is described  by  the  action
(\ref{eq:action}),  where $v=h/a$.   Hence,  we can  use our  previous
results  to define  a vacuum  state for  gravitational waves.   Let us
focus  on a single  comoving $k$-mode  (we drop  in the  following the
subscript  $k$).    It  is  reasonable  to  expect   that  the  vacuum
prescription for gravitons and scalars  are the same.  Hence, we shall
assume that  the quantum state of  the field is the  vacuum defined by
Eq.   (\ref{eq:prescription}) at  the  time of  cutoff  crossing, Eq.  
(\ref{eq:crossing}).   The prescription  (\ref{eq:prescription}) fixes
the  initial conditions  for the  mode functions  $v_\mathrm{in}$, and
hence uniquely determines  the annihilation operators $a_\mathrm{in}$. 
We     shall    call    such     a    vacuum     the    ``in-vacuum'',
$a_\mathrm{in}|\mathrm{in}\rangle=0$.  At an arbitrary later moment of
time $\eta_\mathrm{out}$, an observer  would use the same prescription
(\ref{eq:prescription}) (with  the same  values of the  parameters $X$
and $Y$) to define a  new vacuum $|\mathrm{out}\rangle$.  He would use
a constant time hypersurface to define  a vacuum state at that moment. 
Because  $\theta$ changes  with  time, such  a  prescription yields  a
different set of modes $v_\mathrm{out}$  and hence, a set of different
annihilation  operators  $a_\mathrm{out}$.   Let  us call  the  latter
vacuum  the  out-vacuum,  $a_\mathrm{out}|\mathrm{out}\rangle=0$.   In
general the  in-vacuum and  out-vacuum are different.   In particular,
the in-vacuum  contains particles  of the out-vacuum.   It is  easy to
show that if the out-modes are expressed in terms of the in-modes,
\begin{equation}\label{eq:transformation}
  v_\mathrm{in}=\alpha\, v_\mathrm{out}+\beta\, v_\mathrm{out}^*
\end{equation} 
the                number               of               out-particles
$N_\mathrm{out}=a^\dag_\mathrm{out}a_\mathrm{out}$  contained  in  the
in-vacuum is
\begin{equation}\label{eq:number}
  \langle
  \mathrm{in}|N_\mathrm{out}|\mathrm{in}\rangle=|\beta|^2.
\end{equation} 
This phenomenon is known  as gravitational particle production.  Using
Eq.   (\ref{eq:transformation}) and  the  solution (\ref{eq:solution})
one can easily compute the Bogolubov coefficient $\beta$,
\begin{equation}\label{eq:bogolubov}
\beta=\frac{2}{\pi}(A_\mathrm{out}B_\mathrm{in}
-B_\mathrm{out}A_\mathrm{in}).
\end{equation} 
Here, $A$  and $B$ are the coefficients  (\ref{eq:AandB}) evaluated at
the corresponding  time: at cutoff crossing for  the ``in'' subscript,
and at  $\eta_\mathrm{out}$ for the ``out'' subscript.   Note that the
Bogolubov coefficient  vanishes if $\eta_\mathrm{out}$ is  taken to be
the  time of  cutoff crossing;  the mode  is indeed  in the  vacuum at
cutoff crossing. 

The number of  particles in each $k$-mode can be  used to estimate the
total energy density of  particles produced by the gravitational field
at any arbitrary moment of time $\eta_\mathrm{out}$.  The differential
particle  density  $dn$  is   (we  restore  the  $k$-subindex  in  the
following)
\begin{equation}
  dn=\frac{1}{4\pi^2}\frac{|\beta_k|^2}{a^3}k^2 dk.
\end{equation} 
Since we  are working to  lowest order in  $\theta$ we can  assume the
energy of the graviton to be given by $E=k/a$.  Thus, the differential
energy density of produced particles is
\begin{equation}\label{eq:drho}
  d\rho=\frac{1}{4\pi^2}\frac{|\beta_k|^2}{a^4}k^3 dk.
\end{equation}
The  total energy  density  contributed by  the  created particles  is
obtained by  integrating Eq. (\ref{eq:drho}) over a  suitable range of
modes. At  any given moment  of time $\eta_\mathrm{out}$,  the maximum
value     of     $k$      is     determined     by     the     cutoff,
$k_\mathrm{max}=a(\eta_\mathrm{out})  \Lambda$.   Because  our  vacuum
prescription  (\ref{eq:prescription})  only   makes  sense  for  modes
well-inside the  Hubble radius, we  shall take $k_\mathrm{min}$  to be
determined   by   the   size    of   the   Hubble   radius   at   time
$\eta_\mathrm{out}$,
$k_\mathrm{min}=a(\eta_\mathrm{out})H_\mathrm{out}$.  Substituting Eq.
(\ref{eq:bogolubov}) into Eq.  (\ref{eq:drho}) and integrating between
$k_\mathrm{min}$ and $k_\mathrm{max}$ one obtains to lowest order
\begin{equation}\label{eq:density}
  \rho=\frac{H_\mathrm{out}^4}{16\pi^2}
  \left(\frac{H_\mathrm{out}}{\Lambda}\right)^{-2}
  |X|^2 \cdot 
  I\left(\frac{H_\mathrm{out}}{\Lambda}\right),
\end{equation}
where $I(r)$ is the integral
\begin{eqnarray}
  I(r)=\lefteqn{\int_r^1 \Big\{x^{3-2/p}+x-} \\
  & {}-2 x^{2-1/p}
  \cos\left[\frac{p}{1-p}\frac{\Lambda}{H_\mathrm{out}}
    \left(x^{1/p}-x\right)
  \right]\Big\} dx. \nonumber
\end{eqnarray}
For large  values of $\Lambda/H_\mathrm{out}$ the  oscillatory term in
the  integral averages  out. Then,  for  $p>1/2$ the  integral $I$  is
dominated by the upper limit ($x=1$), i.e. by the high momentum modes,
and the integral is of order  one.  For values of $p$ smaller or equal
$1/2$    the   integral    is   dominated    by   the    lower   limit
($x=H_\mathrm{out}/\Lambda$), i.e.  by the modes close to the horizon.
Because  for those  modes  our vacuum  prescription  breaks down,  and
because  horizon size  modes at  that time  crossed the  cutoff during
inflation, we shall restrict our attention to $p>1/2$ (matter
domination or cosmic acceleration).

One  can  use  Eq.   (\ref{eq:density})  and  the  Friedmann  equation
$H^2=8\pi  \rho_\mathrm{crit}/(3 M_{Pl}^2)$  to compute  the  ratio of
produced particles to the total energy density in a flat universe.  As
an example, we shall consider  our recent past, where the universe has
been mostly  matter dominated ($p=2/3$).  Then,  $I(r)\approx 3/2$ and
we find
\begin{equation}\label{eq:ratio}
\frac{\rho}{\rho_\mathrm{crit}}\approx \frac{1}{4\pi}
\left(\frac{\Lambda}{M_{Pl}}\right)^2|X|^2 .
\end{equation}
Presently, several cosmological  probes constrain the contributions of
the  different universe  constituents to  the critical  energy density
\cite{MAP}.   The  uncertainties  in  those contributions  are  within
$10\%$. Thus,  requiring the energy density of  the produced particles
to be less than $1/10$ of the critical density, we find
\begin{equation}\label{eq:limit}
|X|^2\leq 1.3
\left(\frac{\Lambda}{M_{Pl}}\right)^{-2}.
\end{equation}
If one sets $\Lambda=M_{Pl}$ one finds $|X|\leq 1.1$, which is still
consistent with the Danielsson prescription.  For lower values of the
cutoff, condition (\ref{eq:limit}) is even laxer. Similar limits apply
during inflation, since the inflaton has to remain the dominant energy
component during that epoch.

\section{Summary and Conclusions}
Due to the ambiguity in the choice of a vacuum state during inflation,
we have followed a phenomenological approach to characterize different
possible vacuum  choices. The vacuum  ambiguity has two  sources.  The
first  one is  the  time at  which  the vacuum  is  defined.  We  have
considered two  alternatives: cutoff  crossing and constant  $H$.  The
second source consists on the way a particular vacuum is chosen at the
specified time. We  have parametrized any possible vacuum  by a set of
two  complex  parameters  $X$  and  $Y$,  subject  to  the  constraint
(\ref{eq:constraint}).   These parameters characterize  the departures
from  Minkowski  vacuum due  to  the expansion  to  first  order in  a
properly chosen  expansion parameter: the ratio of  physical length of
the mode to the Hubble radius.

To  first order,  only the  parameter $X$  enters the  predicted power
spectra  generated  during  a  stage of  inflation.   Generically,  if
initial conditions  are chosen at cutoff crossing,  corrections to the
conventional predictions  of inflation  have the form  of oscillations
with    an    amplitude    proportional    to   $H/\Lambda$,    Eq.    
(\ref{eq:corrected-power}).  By  ``generically'' we mean  that the set
of $X$ values  that yield no linear corrections is  of zero measure in
parameter space.  Nevertheless, for the particular vacuum choices that
have been traditionally discussed  in the literature (adiabatic vacuum
and Hamiltonian  diagonalization), corrections are at  most quadratic. 
If initial conditions are specified  at a constant time, the amplitude
of the oscillations  is proportional to a large  power of $H/\Lambda$,
i.e.  it is highly suppressed.

Since theoretical  arguments might  not be able  to constrain  $X$ and
$Y$,  we have  attempted to  set limits  on their  values  by studying
graviton creation  in an expanding  universe. By requiring  the energy
density of the produced gravitons to  be less than ten per cent of the
critical energy density  today we obtain a constraint  on the value of
$X$, Eq.   (\ref{eq:limit}).  Even if the  cutoff is chosen  to be the
Planck  energy,  the  Danielsson  prescription  is  still  allowed  by
observations.   For  lower values  of  the  cutoff, our  observational
constraint is not significant.

We conclude  that, at the level  of our analysis, the  issue about the
choice of vacuum is observational, rather than theoretical.  This does
not  mean however  that inflation  is not  predictive. All  the vacuum
choices we  have discussed yield the  same results at  zeroth order in
our   expansion   parameter.    The   first  order   corrections   are
``Planck''-suppressed,  i.e.  they  are proportional  to  $H/\Lambda$. 
Thus, these  corrections are always  small, albeit in some  cases they
might be big enough to be detectable.

\acknowledgments We  thank Stefan Hollands,  Wayne Hu and  Finn Larsen
for  useful conversations, and  Sean Carroll,  Jim Chisolm  and Takemi
Okamoto for valuable  comments on a draft of  this manuscript. We also
thank Leonard  Parker for clarifying  correspondence on the  nature of
the  adiabatic vacuum.  The  discussion of  the Hamiltonian  vacuum is
based on unpublished  lecture notes by V.  Mukhanov.   CAP and EL were
supported by the U.S.  DoE grant DE-FG02-90ER40560.

\appendix

\section{Vacuum prescriptions}
We discuss in the following  most of the different vacuum choices that
have  been proposed  in the  literature.  They  are based  on imposing
different conditions  on the vacuum state.   Up to what  we shall call
the ``conventional'' choice, they all share the same ambiguity related
to the time  where initial conditions for the  mode functions shall be
set.

\subsection{Adiabatic vacua}
The adiabatic  vacuum \cite{Parker,adiabatic}  has been claimed  to be
the best available notion of vacuum  in a spacetime that has no static
regions \cite{BirrellDavies} or  any further additional symmetry.  The
notion of  adiabatic vacuum  is slightly technical,  and has  been the
origin of some confusion in  the literature.  We hence review here the
presentation of \cite{BirrellDavies}.
 
The  first ingredient  of  the adiabatic  vacuum  prescription is  the
exact, formal  ``positive frequency'' WKB solution of  the equation of
motion (\ref{eq:motion}),
\begin{equation}\label{eq:WKB}
  v_k=\frac{1}{\sqrt{W_k}}
  \exp\left(-i\int^\eta W_k(\tilde{\eta})\,d\tilde{\eta}\right),
\end{equation}
where $W_k$ is implicitly defined by the equation
\begin{equation}\label{eq:W}
W_k^2=\omega_k^2-\frac{1}{2}\left(\frac{W_k''}{W_k}-\frac{3}{2}
\frac{W_k'^2}{W_k^2}\right),
\end{equation}
and $\omega_k$ is given by Eq. (\ref{eq:frequency}),
\begin{equation}\label{eq:power-frequency}
  \omega_k^2=k^2\left(1-\frac{2p-1}{p}\theta^2\right).
\end{equation}
The  reader  can  easily   verify  that  the  function  (\ref{eq:WKB})
satisfies the normalization condition (\ref{eq:normalization}).

In  the limit where  the expansion  of the  universe is  ``slow'', one
expects  an  expansion in  the  number of  derivatives  to  be a  good
approximation to  solve for  $W_k$. To perform  such an  expansion one
formally replaces $\eta$  by $\eta T$ and expands  all quantities in a
power series in $T^{-1}$.  To $n$th adiabatic order, only powers up to
$T^{-n}$ are kept.  Such an  expansion is the second ingredient of the
adiabatic prescription.  It can be  easily verified that to lowest and
first adiabatic  order $\omega_k=k$ and  $W_k=\omega_k=k$.  Nontrivial
time-dependent corrections do not show up until 2nd order.

If one  substitutes the $n$th  order adiabatic approximation  to $W_k$
into  (\ref{eq:WKB}),  one  obtains  the $n$th  order  adiabatic  mode
$v_k^{(n)}$.  Note however that  this $n$th order approximation is not
an exact solution of the mode equation (\ref{eq:motion}) anymore.  The
third  step of  the adiabatic  prescription consists  of  matching the
$n$th  order adiabatic  approximation  $v_k^{(n)}$ to  the exact  mode
solutions of Eq. (\ref{eq:motion})  at an arbitrary time $\eta_0$.  In
other words,  the $n$th adiabatic  approximation is used  to prescribe
the values of $v_k$ and $v_k'$ at time $\eta_0$,
\begin{equation}\label{eq:adiabatic-matching}
  v_k(\eta_0)=v_k^{(n)}(\eta_0), \quad
  v_k'(\eta_0)={v_k'}^{(n)}(\eta_0).
\end{equation}
Because the  WKB solution  has been expanded  only to  $n$th adiabatic
order, the  matching of the  two functions has  to be carried  only to
that same order, i.e. only terms  containing up to $n$ powers of $1/T$
have  to be  matched.   Conditions (\ref{eq:adiabatic-matching})  then
uniquely define the mode function  $v_k$. By construction, $v_k$ is an
exact  solution  of Eq.   (\ref{eq:motion}),  with initial  conditions
determined by  the ``distorted''  adiabatic modes $v_k^{(n)}$  at time
$\eta_0$.

If  one  carries  over   the  program  described  above,  one  obtains
expressions for  $A_k$ and $B_k$ as  a power series  in $1/(T\eta_0)$. 
To $n$th  adiabatic order only  terms up to order  $1/(T\eta_0)^n$ are
significant.   Hence, if we  are interested  in determining  $A_k$ and
$B_k$  to  first  order  in $\theta_0\propto  1/(T\eta_0$),  it  hence
suffices  to consider  the 1st  order adiabatic  vacuum;  higher order
adiabatic vacua only yield higher order corrections.  Because to first
adiabatic      order     $W_k=k$,      the      matching     condition
(\ref{eq:adiabatic-matching}) to  that same adiabatic  order reads, up
to an irrelevant common phase,
\begin{equation}
v_k(\eta_0)= \frac{1}{\sqrt{k}},\quad
v'_k(\eta_0)=-i \sqrt{k}.
\end{equation}
By comparison with Eqs. (\ref{eq:prescription}) we thus find
\begin{equation}
  Y=0,\quad X=0.
\end{equation}  

\subsection{The conventional choice}
Conventionally,  the  vacuum is  chosen  by  requiring  that the  mode
functions   $v_k$  reduce  to   the  Minkowski   ones  in   the  limit
$\eta\to-\infty$,
\begin{equation}\label{eq:conventional}
v_k(-\infty)= \frac{1}{\sqrt{k}},\quad
v'_k(-\infty)=-i \sqrt{k}.
\end{equation}

Using  the expansion  (\ref{eq:expansion}) one  can directly  find the
values of the coefficients $A_k$ and $B_k$ for which the mode function
(\ref{eq:solution}) satisfies Eqs. (\ref{eq:conventional}),
\begin{equation}
  A_k=\sqrt{\frac{\pi}{2}}, \,\quad B_k=0.
\end{equation}
Once these coefficients  are known one can then  determine $X$ and $Y$
by comparing them Eqs. (\ref{eq:AandB}),
\begin{equation}\label{eq:standard-xi}
  X=0,\quad
  Y=i\, \frac{1-2p}{1-p}.
\end{equation}
The  conventional vacuum  is  sometimes called  the adiabatic  vacuum,
since it agrees with the adiabatic  vacuum (of any order) in the limit
$\eta\to-\infty$.   One   should  bear   in  mind  however   that  the
conventional vacuum  is only  \emph{a particular} adiabatic  vacuum in
the infinite set  of adiabatic vacua parametrized by  $\eta_0$. Let us
note in  addition that  all the vacuum  prescriptions we  discuss here
also yield the adiabatic vacuum in the limit $\eta_0\to -\infty$.

\subsection{Hamiltonian diagonalization}

The Hamiltonian diagonalization vacuum\footnote{During the preparation
  of this  work the  preprint \cite{BoGiVe} appeared,  which discusses
  how  the  choice of  different  Hamiltonians  affects  the state  of
  minimal  energy.}    is  the  state  that   diagonalizes  the  field
Hamiltonian (\ref{eq:Hamiltonian}) at a  given moment of time $\eta_0$
\cite{MuFeBr,BirrellDavies}.  Alternatively, it  can be shown that the
Hamiltonian  diagonalization vacuum  is the  state that  minimizes the
energy  $E=\langle 0|\mathcal{H}|0\rangle$  at $\eta_0$.   It directly
follows   from   the   Hamiltonian  (\ref{eq:Hamiltonian})   and   the
normalization    condition     (\ref{eq:normalization})    that    the
diagonalization   prescription  amounts   to   imposing  the   initial
conditions
\begin{equation}\label{eq:diagonalization}
  v_k(\eta_0)=\frac{1}{\sqrt{\omega_k(\eta_0)}},\quad
  v'_k(\eta_0)=-i \sqrt{\omega_k(\eta_0)},
\end{equation}
where $\omega_k$  is the frequency (\ref{eq:frequency}).  Note that it
is  only possible  to  diagonalize the  Hamiltonian  for a  particular
subset of field modes.

The  squared  frequency  (\ref{eq:power-frequency})  is  quadratic  in
$\eta_0$.   Hence   the  expansion  of   the  square  roots  in   Eq.  
(\ref{eq:diagonalization}) in  a power  series in $\theta_0$  does not
contain linear  terms ($\propto \theta_0$).  Therefore, we immediately
find
\begin{equation}
  X=0,\quad Y=0.
\end{equation}
In the absence  of a cutoff, Hamiltonian diagonalization  has has been
criticized   on   the  grounds   of   excessive  particle   production
\cite{BirrellDavies}.  Note  that the Hamiltonian  diagonalization and
1st order adiabatic vacua are the same.

\subsection{Danielsson prescription}

According to  Danielsson \cite{Danielsson, EGKS}, the  vacuum shall be
chosen by imposing the following condition on the mode functions,
\begin{equation}\label{eq:Danielsson}
\left(\frac{v_k}{a}\right)'(\eta_0)=-i k \,\frac{v_k}{a}(\eta_0).
\end{equation}  
Using  the  normalization   condition  (\ref{eq:normalization}),  Eq.  
(\ref{eq:Danielsson}) implies the following  relations (up to a common
irrelevant phase)
\begin{equation}\label{eq:Danielsson-v}
v_k(\eta_0)=\frac{1}{\sqrt{k}},\quad 
v_k'(\eta_0)=-i\sqrt{k}\left(1+i\theta_0\right).
\end{equation}
Therefore,   by comparing  Eqs.    (\ref{eq:prescription})   with  Eqs.   
(\ref{eq:Danielsson-v}) it follows
\begin{equation}
X=-i,\quad Y=i. 
\end{equation}

The   Danielsson   vacuum   is   a  state   of   minimal   uncertainty
\cite{StarobinskyPolarski}, but not the only one.  It is not the state
of  minimal  energy  either,  in  the  sense  that  it  minimizes  the
Hamiltonian   (\ref{eq:Hamiltonian})   (see   also   \cite{BoGiVe}).   
Nevertheless, at  this stage the  prescription (\ref{eq:Danielsson-v})
is as valid as any other.


\begin{thebibliography}{99}
\bibitem{Linde} A. D.  Linde,
\textit{Inflationary Cosmology and Particle Physics}, Harwood (1990).

\bibitem{MuFeBr}
V.~F.~Mukhanov, H.~A.~Feldman and R.~H.~Brandenberger,
\textit{Theory Of Cosmological Perturbations},
Phys.\ Rept.\  {\bf 215}, 203 (1992).

\bibitem{BirrellDavies}
N.~D.~Birrell and P.~C.~Davies,
\textit{Quantum Fields In Curved Space}, Cambridge University Press (1982).

\bibitem{vacua}
U.~H.~Danielsson,
\textit{Inflation, holography and the choice of vacuum in de Sitter space},
JHEP {\bf 0207}, 040 (2002); \texttt{hep-th/0205227}.

T.~Banks and L.~Mannelli, \textit{de Sitter vacua, renormalization and
  locality}, \texttt{hep-th/0209113}.

M.~B.~Einhorn and F.~Larsen,
\textit{Interacting quantum field theory in de Sitter vacua},
Phys.\ Rev.\ D {\bf 67}, 024001 (2003);
\texttt{hep-th/0209159}.

N.~Kaloper, M.~Kleban, A.~Lawrence, S.~Shenker and L.~Susskind,
\textit{Initial conditions for inflation},
JHEP {\bf 0211}, 037 (2002);
\texttt{hep-th/0209231}.

U.~H.~Danielsson,
\textit{On the consistency of de Sitter vacua},
JHEP {\bf 0212}, 025 (2002);
\texttt{hep-th/0210058}.

K.~Goldstein and D.~A.~Lowe,
\textit{A note on alpha-vacua and interacting field theory in de Sitter space}
, \texttt{hep-th/0302050}.

\bibitem{DeruelleMukhanov}
N.~Deruelle and V.~F.~Mukhanov,
\textit{On matching conditions for cosmological perturbations},
Phys.\ Rev.\ D {\bf 52}, 5549 (1995);
\texttt{gr-qc/9503050}.

\bibitem{Wald} R.~M.~Wald, \textit{Quantum Field Theory In Curved Space-Time
  And Black Hole Thermodynamics}, University of Chicago Press (1984).

R.~M.~Wald,
\textit{Quantum field theory in curved spacetime},
\texttt{gr-qc/9509057}.

\bibitem{Brandenberger}
R.~H.~Brandenberger,
\textit{Inflationary cosmology: Progress and problems},
\texttt{hep-ph/9910410}.

\bibitem{Jacobson}
T.~Jacobson,
\textit{Black hole radiation in the presence of a short distance cutoff},
Phys.\ Rev.\ D {\bf 48}, 728 (1993); \texttt{hep-th/9303103}.

\bibitem{Danielsson}
U.~H.~Danielsson,
\textit{A note on inflation and transplanckian physics},
Phys.\ Rev.\ D {\bf 66}, 023511 (2002); \texttt{hep-th/0203198}.

\bibitem{Brandenberger2}     R.~H.~Brandenberger,    \textit{Trans-Planckian
  physics and inflationary cosmology}, \texttt{hep-th/0210186}.

\bibitem{BergstromDanielsson}
L.~Bergstrom and U.~H.~Danielsson,
\textit{Can MAP and Planck map Planck physics?},
JHEP {\bf 0212}, 038 (2002);
\texttt{hep-th/0211006}.

\bibitem{LimOkamoto}
E. Lim and T. Okamoto, in preparation.

\bibitem{KKLS} N.~Kaloper, M.~Kleban, A.~E.~Lawrence and S.~Shenker,
  \textit{Signatures of short distance physics in the cosmic microwave
    background}, Phys.\ Rev.\ D {\bf 66}, 123510 (2002);
  \texttt{hep-th/0201158}.

\bibitem{Starobinsky}  A.~A.~Starobinsky,  \textit{Robustness  of  the
    inflationary  perturbation spectrum  to  trans-Planckian physics},
  JETP  Lett.\  {\bf   73},  371  (2001);  \texttt{astro-ph/0104043}.  
  A.~A.~Starobinsky    and    I.~I.~Tkachev,   \textit{Trans-Planckian
    particle creation in cosmology and ultra-high energy cosmic rays},
  JETP Lett.\ {\bf 76}, 235 (2002); \texttt{astro-ph/0207572}.

\bibitem{GarrigaMukhanov}      J.~Garriga      and     V.~F.~Mukhanov,
 \textit{Perturbations  in k-inflation},  Phys.\ Lett.\  B {\bf  458}, 219
  (1999); \texttt{hep-th/9904176}.

\bibitem{GasperiniVeneziano}
M.~Gasperini and G.~Veneziano,
\textit{Pre - big bang in string cosmology},
Astropart.\ Phys.\  {\bf 1}, 317 (1993);
\texttt{hep-th/9211021}.

\bibitem{EpCy} J.~Khoury, B.~A.~Ovrut, P.~J.~Steinhardt and N.~Turok,
  \textit{The ekpyrotic universe: Colliding branes and the origin of
    the hot big bang}, Phys.\ Rev.\ D {\bf 64}, 123522 (2001);
  \texttt{hep-th/0103239}.

P.~J.~Steinhardt  and   N.~Turok,  \textit{Cosmic  evolution   in  a  cyclic
universe},    Phys.\    Rev.\    D    {\bf   65},    126003    (2002);
\texttt{hep-th/0111098}.

\bibitem{AADD}
I.~Antoniadis, N.~Arkani-Hamed, S.~Dimopoulos and G.~R.~Dvali,
\textit{New dimensions at a millimeter to a Fermi and superstrings at a TeV},
Phys.\ Lett.\ B {\bf 436}, 257 (1998);
\texttt{hep-ph/9804398}.

\bibitem{AbramowitzStegun}M.     Abramowitz     and     I.     Stegun,
  \textit{Handbook  of  Mathematical  Functions},  Dover  Publications
  (1965).

\bibitem{MAP}  D.~N.~Spergel  {\it  et  al.}, \textit{First  Year  Wilkinson
  Microwave  Anisotropy Probe  (WMAP)  Observations: Determination  of
  Cosmological   Parameters}, \texttt{astro-ph/0302209}.

\bibitem{InflMAP}  V.~Barger, H.~S.~Lee and  D.~Marfatia, \textit{WMAP
    and     inflation},     \texttt{hep-ph/0302150}.     S.~L.~Bridle,
  A.~M.~Lewis, J.~Weller and G.~Efstathiou, \textit{Reconstructing the
    primordial    power    spectrum},    \texttt{astro-ph/0302306}.    
  C.~R.~Contaldi, H.~Hoekstra and A.~Lewis, \textit{Joint CMB and Weak
    Lensing Analysis: Physically Motivated Constraints on Cosmological
    Parameters}, \texttt{astro-ph/0302435}.

\bibitem{EGKS}
R.~Easther, B.~R.~Greene, W.~H.~Kinney and G.~Shiu,
\textit{A generic estimate of trans-Planckian modifications to the primordial  power spectrum in inflation},
Phys.\ Rev.\ D {\bf 66}, 023518 (2002);
\texttt{hep-th/0204129}.

\bibitem{NiPaCa}    J.~C.~Niemeyer,    R.~Parentani   and    D.~Campo,
  \textit{Minimal modifications of  the primordial power spectrum from
    an  adiabatic short  distance cutoff},  Phys.\ Rev.\  D  {\bf 66},
  083510 (2002); \texttt{hep-th/0206149}.
  
\bibitem{Parker}
L.~Parker,
\textit{Quantized Fields And Particle Creation In Expanding Universes. 1},
Phys.\ Rev.\  {\bf 183}, 1057 (1969).

\bibitem{adiabatic}  L.~Parker  and  S.~A.~Fulling,  \textit{Adiabatic
    Regularization Of The Energy  Momentum Tensor Of A Quantized Field
    In Homogeneous Spaces}, Phys.\ Rev.\ D {\bf 9}, 341 (1974).

\bibitem{BoGiVe}
V. Bozza, M. Giovannini and G. Veneziano, \textit{Cosmological Perturbations
from a New-Physics Hypersurface},
\texttt{hep-th/0302184}.

\bibitem{StarobinskyPolarski}
 D.~Polarski and A.~A.~Starobinsky,
\textit{Semiclassicality and decoherence of cosmological perturbations},
Class.\ Quant.\ Grav.\  {\bf 13}, 377 (1996);
\texttt{gr-qc/9504030}.

\end{thebibliography}
\end{document}